\documentstyle[12pt,titlepage,epsfig,floatfig]{article}
\begin{document}
\newcommand{\gapproxeq}{\lower 
.7ex\hbox{$\;\stackrel{\textstyle >}{\sim}\;$}}
\baselineskip=0.163in

\initfloatingfigs

\begin{flushleft}
\large
{SAGA-HE-104-95 
\hfill June 30, 1996}  \\
\end{flushleft}
 
\vspace{3.0cm}
 
\begin{center}
 
\LARGE{{\bf $\bf Q^2$ evolution studies of}} \\

\vspace{0.3cm}

\LARGE{{\bf nuclear structure function $\bf F_2$ at HERA}} \\
 
\vspace{2.0cm}
 
\Large
{S. Kumano and M. Miyama $^*$}   \\
 
\vspace{1.0cm}
  
\Large
{Department of Physics}         \\
 
\vspace{0.1cm}
 
\Large
{Saga University}      \\
 
\vspace{0.1cm}

\Large
{Saga 840, Japan} \\

\vspace{2.2cm}
 
{contribution to ``Future Physics at HERA''} \\
 
\end{center}
 
\vspace{1.3cm}
\vfill
 
\noindent
{\rule{6.cm}{0.1mm}} \\
 
\vspace{-0.4cm}

\noindent
\vspace{-0.2cm}
\normalsize
{* Email: kumanos or 96td25@cc.saga-u.ac.jp}  \\

\vspace{-0.2cm}
\noindent
\normalsize
{Research activities are listed 
 at ftp://ftp.cc.saga-u.ac.jp/pub/paper/riko/quantum1} \\
\vspace{-0.6cm}

\noindent
{or at 
http://www.cc.saga-u.ac.jp/saga-u/riko/physics/quantum1/structure.html.} \\

\vfill\eject

\pagestyle{plain}
\begin{center}
 
\large
{$Q^2$ evolution studies of nuclear structure function $F_2$ at HERA} \\

\vspace{0.5cm}
 
{S. Kumano and M. Miyama} \\

{Department of Physics, Saga University, Saga 840, Japan}  \\
 
\end{center}

\vspace{0.6cm}


Nuclear modification of the structure function $F_2$ has 
been an interesting topic since the discovery of the EMC effect in 1983.
Although most studies discuss $x$ dependence of the modification,
$Q^2$ dependence becomes increasingly interesting.
It is because the NMC measured $Q^2$ variations 
of the ratio $F_2^A/F_2^D$ \cite{NMC}. 
Furthermore, it is found recently that 
there exist significant differences between tin and carbon $Q^2$
variations, $\partial [F_2^{Sn}/F_2^C]/ \partial [\ln Q^2]\ne 0$ \cite{NMC}.
However, the NMC data are taken in the limited small $Q^2$ range
at small $x$, so that they are not sufficient to test 
nuclear $Q^2$ evolution.
The $Q^2$ dependence is important for understanding perturbative QCD
in nuclear environment, and the future HERA nuclear program can
make important contributions to this interesting topic.

The $Q^2$ dependence of structure functions can be calculated
by using the DGLAP equations. 
They have been successful in describing many experimental data.
However, as it becomes possible to reach the small $x$ region
by high-energy accelerators, it is necessary to investigate
the details of small $x$ physics. 
The longitudinal localization size of a parton exceeds
the average nucleon separation in a nucleus
in the small $x$ region ($x<0.1$).
It means that partons in different nucleons could interact 
in the nucleus, and the interaction is called parton recombination (PR).
This mechanism is used for explaining nuclear shadowing.
There are a number of studies on the recombinations.
Among them, we employ the evolution equations proposed by
Mueller and Qiu. They investigated gluon-gluon recombination 
effects on the evolution.
The DGLAP and PR  evolution equations are given by
(see Ref. \cite{MK} for the details)
$$
{\partial \over {\partial t}} \ q_i \left({x,t}\right)\
=\ \int_{x}^{1}{dy \over y}\ 
\left[\ \sum_j P_{q_{i} q_{j}}\left({{x \over y}}\right)\ 
           q_j \left({y,t}\right)\ 
+\  P_{qg}\left({{x \over y}}\right)\ 
g\left({y,t}\right)\ \right]
$$
\vspace{-0.3cm}
$$ 
\ \ \ \ \ \ \ \ \ \ \ \ \ \ \ \ \ \ \ \ \ \ \ \ \ \ \ 
\ \ \ \ \ \ \ \ \ \ \ \ \ \ \ \ \ \ \ \ \ \ \ \ \ \ \ 
+ \ \left( recombination\ terms\ \propto \ 
{{\alpha_s A^{1/3}} \over {Q^2}} \right) \ 
\ ,
\eqno{(1a)}
$$
$$
{\partial \over {\partial t}} \ g\left({x,t}\right)\ 
=\ \int_{x}^{1}{dy \over y}\ 
\left[\ \sum_j P_{gq_j}\left({{x \over y}}\right)\ 
q_j \left({y,t}\right)\ 
+\ P_{gg}\left({{x \over y}}\right)\ 
g\left({y,t}\right)\ \right]
$$
\vspace{-0.3cm}
$$
\ \ \ \ \ \ \ \ \ \ \ \ \ \ \ \ \ \ \ \ \ \ \ \ \ \ \ 
\ \ \ \ \ \ \ \ \ \ \ \ \ \ \ \ \ \ \ \ \ \ \ \ \ \ \ 
+ \ \left( recombination\ terms\ \propto \ 
{{\alpha_s A^{1/3}} \over {Q^2}} \right) \ 
\ ,
\eqno{(1b)}
$$
where the variable $t$ is defined by
$t = -(2/\beta_0) \ln [\alpha_s(Q^2)/\alpha_s(Q_0^2)]$.
In the PR evolution case, there is an extra evolution
equation for a higher-dimensional gluon distribution.
The first two terms in Eqs. (1a) and (1b) describe the process
that a parton $p_j$ with the nucleon's momentum fraction $y$ 
splits into a parton $p_i$ with the momentum fraction $x$
and another parton.
The splitting function $P_{p_i p_j}(z)$ determines 
the probability that such a splitting process occurs and
the $p_j$-parton momentum is reduced by the fraction $z$. 

Although the DGLAP equations are well tested by 
various experimental data, the PR equations are not well established yet.
An interesting problem is possible nuclear dependence in
the $Q^2$ evolution.
There are two possible sources for the nuclear dependence in 
the evolution equations. One is the input parton distributions, 
and another is the recombination effects.
The modification of the input $x$-distributions in a nucleus
affects the $Q^2$ evolution through splitting functions.
The recombination contributions enter into the evolution
equations as additional higher-twist effects.

In studying the $Q^2$ evolution, it is very important to have
correct input distributions. Fortunately, there are many data on the
$x$ dependence of $F_2^A/F_2^D$, so that we could have reasonable
nuclear input distributions. 
We employ a hybrid parton model with recombination and $Q^2$ rescaling 
mechanisms in Ref. \cite{KMU}. However, it does not matter 
in the $Q^2$ evolution studies what kind of model is used
if it can explain the experimental
$x$ dependence of $F_2^A/F_2^D$.
In the hybrid model, we first calculate $Q^2$ rescaled
valence-quark distributions at $Q_0^2$. Sea-quark and gluon
distributions are simply modified by a constant mount so as to satisfy
the momentum conservation. Then, obtained distributions are
used as input distributions for calculating the recombination effects.
In this way, nuclear parton distribution with the rescaling and 
recombination effects are obtained at $Q_0^2$.
Because the recombinations are higher-twist effects,
final distributions are very sensitive to the choice of
$Q_0^2$. It is fixed so that obtained shadowing agrees with
the NMC ratios $F_2^{Ca}/F_2^D$ at small $x$. 
In the following, we discuss two topics on the $Q^2$ evolution.
The first is $Q^2$ variation of $F_2^A/F_2^D$ \cite{MK,KMU} 
and the second is
$\partial [F_2^{Sn}/F_2^C]/ \partial [\ln Q^2]$ \cite{KM}.

\begin{floatingfigure}{5.6cm}
   \begin{center}
      \mbox{\epsfig{file=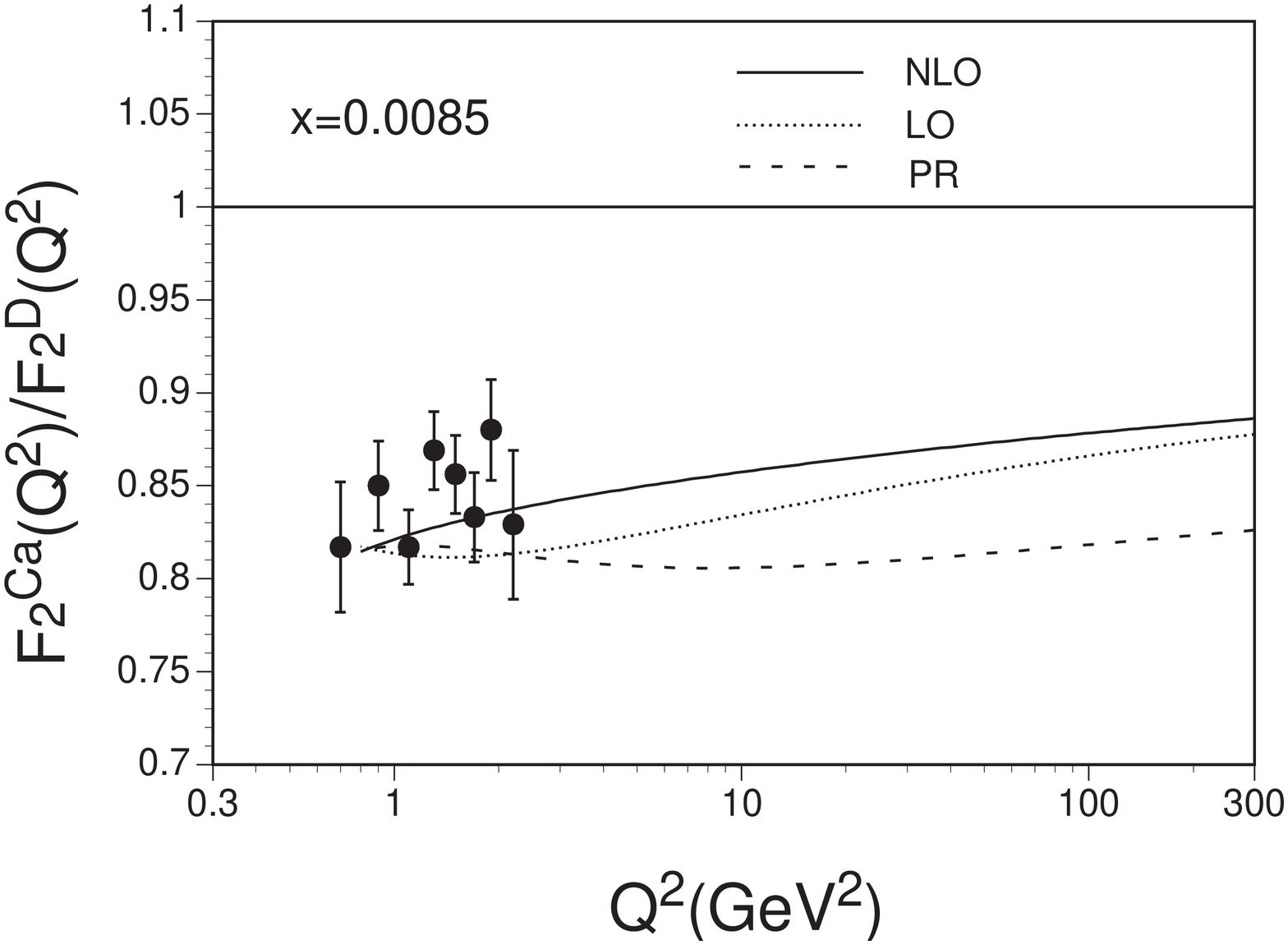,width=5.0cm}}
   \end{center}
 \vspace{-0.8cm}
\caption{\footnotesize $Q^2$ variation of $F_2^{Ca}/F_2^D$. }
\label{fig:fig1}
\end{floatingfigure}
\quad
We compare calculated evolution results with the NMC data in 
Fig. \ref{fig:fig1} at $x$=0.0085 \cite{MK}.
The initial distributions at $Q_0^2$=0.8 GeV$^2$
in the nucleon and the calcium nucleus are taken from Ref. \cite{KMU}.
In Fig. \ref{fig:fig1}, the dotted, solid, and dashed curves are obtained
in the leading-order (LO) DGLAP, next-to-leading-order (NLO) DGLAP,
and NLO evolution equations with parton-recombination contributions
respectively ($\Lambda$=0.2 GeV and $N_f$=3).
As shown in the figure, NLO and recombination
contributions to the ratio
are conspicuous at such a small $x$.
If we evolve $F_2$ from $Q_0^2$=0.8 GeV$^2$,
the recombination effects are larger than the NLO ones.
It is interesting to find such large recombination
contributions in Fig. 1. 
However, the recombination cannot be tested at this stage
because we do not have
the data in the wide $Q^2$ region at small $x$.
The future HERA nuclear program should be able to study 
the large $Q^2$ region, so that the parton recombination mechanism
could be tested.

Next, $Q^2$ evolution differences in various nuclei
could also be investigated at HERA.
There are significant differences between tin and carbon $Q^2$
variations according to recent NMC analysis.
It is the first indication of nuclear effects on the $Q^2$ evolution
of $F_2$. The phenomena are worth investigating theoretically.

\begin{floatingfigure}{5.6cm}
   \begin{center}
      \mbox{\epsfig{file=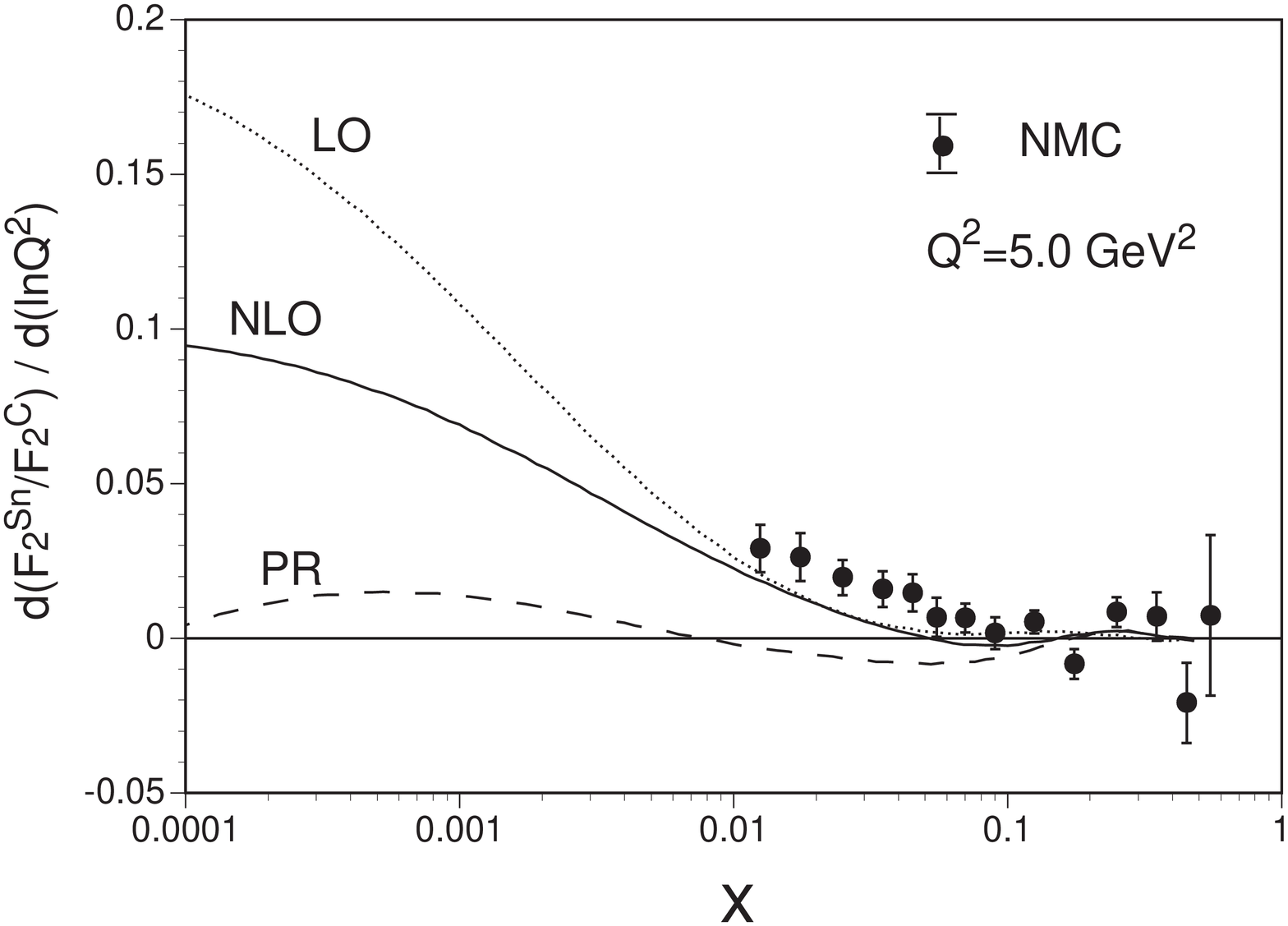,width=5.0cm}}
   \end{center}
 \vspace{-0.8cm}
\caption{\footnotesize Nuclear dependence in $Q^2$ evolution of $F_2$.}
\label{fig:fig2}
\end{floatingfigure}
\quad
The $Q^2$ evolution of the structure functions $F_2$
in tin and carbon nuclei is investigated in Ref. \cite{KM}.
As the input distributions, we employ those in Ref. \cite{KMU}.
$F_2$ is evolved by using LO DGLAP, NLO DGLAP, and PR equations
with the help of a computer program in Ref. \cite{MK}.
Calculated results for $\partial [F_2^{Sn}/F_2^C]/ \partial [\ln Q^2]$
at $Q^2$=5 GeV$^2$ are compared with the NMC data.
The DGLAP evolution curves agree roughly with the experimental
tendency, but the PR results are significantly different from the data.
However, it does not mean that 
the recombination mechanism should be ruled out because there exists
an unknown parameter $K_{HT}$ associated with
the higher-dimensional gluon distribution in the recombination.
In order to discuss the validity of the PR evolution,
the constant $K_{HT}$ must be evaluated theoretically.

In this way, the NMC experimental result 
$\partial [F_2^{Sn}/F_2^C]/ \partial [\ln Q^2]\ne 0$ could be
essentially understood by the difference of parton distributions
in the tin and carbon nuclei together with the ordinary
DGLAP evolution equations.
However, we find an interesting indication that ``large" higher-twist
effects on the $Q^2$ evolution could be ruled out.
As shown in Fig. \ref{fig:fig2}, there are large differences
among three evolution results at small $x$ ($\approx 10^{-4}$).
The future HERA program can study nuclear dependence
of the $Q^2$ evolution ($\partial [F_2^A/F_2^D]/ \partial [\ln Q^2]$)
in this small $x$ region, and it provides
us crucial information on recombination effects and on higher-order
$\alpha_s$ effects.

\vspace{0.4cm}
\noindent
{\bf Acknowledgment} 
\vspace{0.18cm}

This research was partly supported by the Grant-in-Aid for
Scientific Research from the Japanese Ministry of Education,
Science, and Culture under the contract number 06640406.

\vspace{-0.3cm}


\end{document}